\documentclass[aps,pre,twocolumn,superscriptaddress,balancelastpage,showpacs,letterpaper,nofootinbib]{revtex4-2}
\usepackage{graphicx}
\usepackage{xcolor}
\usepackage{amsmath}
\usepackage{amssymb}
\usepackage[color=yellow]{todonotes}

\definecolor{darkblue}{rgb}{0,0,0.6}
\definecolor{darkred}{rgb}{0.6,0,0}
\usepackage[colorlinks=true,urlcolor=darkblue,citecolor=darkblue, linkcolor=darkred,hyperfootnotes=false]{hyperref}
\graphicspath{ {figures/} }

\newcommand{\dd}{\mathrm{d}}
\newcommand{\EE}{\boldsymbol{E}}

\newcommand{\ff}{\boldsymbol{f}}
\newcommand{\jj}{\boldsymbol{j}}
\newcommand{\JJ}{\boldsymbol{J}}
\newcommand{\kk}{\boldsymbol{k}}
\newcommand{\qq}{\boldsymbol{q}}
\newcommand{\mcO}{\mathcal{O}}

\newcommand{\uu}{\boldsymbol{u}}

\newcommand{\rr}{\boldsymbol{r}}

\newcommand{\zzeta}{\boldsymbol{\zeta}}

\newcommand{\ind}[1]{_{\mathrm{#1}}}
\newcommand{\lD}{\lambda\ind{D}}
\newcommand{\lB}{l\ind{B}}

\newcommand{\D}{{\rm d}}
\newcommand{\be}{\begin{equation}}
\newcommand{\ee}{\end{equation}}
\newcommand{\bea}{\begin{eqnarray}}
\newcommand{\eea}{\end{eqnarray}}

\begin{document}

\title{Frequency-Dependent Conductivity of Concentrated Electrolytes:\\
A Stochastic Density Functional Theory}

\author{Haggai Bonneau}
\email{haggai.bonneau@mail.huji.ac.il}
\affiliation{School of Physics and Astronomy, Tel Aviv University, Ramat Aviv 69978, Tel Aviv, Israel} 
\author{Yael Avni}
\email{yael.avni10@gmail.com}
\affiliation{University of Chicago, James Franck Institute, 929 E 57th Street, Chicago, IL 60637}
\author{David Andelman}
\email{andelman@post.tau.ac.il}
\affiliation{School of Physics and Astronomy, Tel Aviv University, Ramat Aviv 69978, Tel Aviv, Israel}
\author{Henri Orland}
\email{henri.orland@ipht.fr}
\affiliation{Institut de Physique Th\'eorique, Universit\'e de Paris-Saclay, CEA, CNRS, F-91191 Gif-sur-Yvette Cedex, France}



\begin{abstract}
The response of ionic solutions to time-varying electric fields, quantified by a frequency-dependent conductivity, is essential 
in many electrochemical applications. Yet, it constitutes a challenging problem due to the combined effect of Coulombic interactions, 
hydrodynamics, and thermal fluctuations. Here, we study the frequency-dependent conductivity of ionic solutions using a stochastic density functional theory. 
In the limit of small concentrations, we recover the classical Debye and Falkenhagen (DF) result, predicting an increase in conductivity with field frequency. 
At higher concentrations, we use a modified Coulomb interaction potential that accounts for the hard-core repulsion between the ions, which was recently 
employed in the zero-frequency case. Consequently, we extend the DF result to concentrated electrolytes. We discuss experimental and numerical studies 
and the complexity of observing the DF effect in such setups. 
\end{abstract}
\maketitle

\section{Introduction}
The transport properties of electrolyte solutions have been the subject of long-standing fundamental and applied research. Ion dynamics 
in a solvent is essential in many physical and chemical processes~\cite{wilson2020diffusiophoresis, bazant2010induced,bazant2004,wall2010history}. 
Furthermore, diverse technological applications of ion transport range from electrochemical energy storage in batteries to the flow of 
ions and charge molecules through biological cell membranes ({\em e.g.,} ion channels and 
pumps)~\cite{simon2020perspectives,Robin2021Modeling,mceldrew2020theory,benaglia2021tip,gebbie2013ionic,smith2016electrostatic}.
 
Debye and H\"uckel were the first to propose a theory of electrolytic conductivity that included ion-ion interactions~\cite{Debye1923}. 
Their approach involved the concept of an ionic cloud, wherein each ion is assumed to be surrounded by a smeared ionic distribution 
of net opposite charge that gets distorted by the central ion's movement. A few years later, Onsager refined the Debye-H\"uckel theory 
to account consistently for central-ion diffusion. This resulted in the century-old Debye-H\"uckel-Onsager (DHO) theory for DC (zero frequency) 
conductivity of electrolytes~\cite{onsager1926, Onsager1957Wien}.

\begin{figure}
	\centering
	\includegraphics[width=0.5\linewidth]{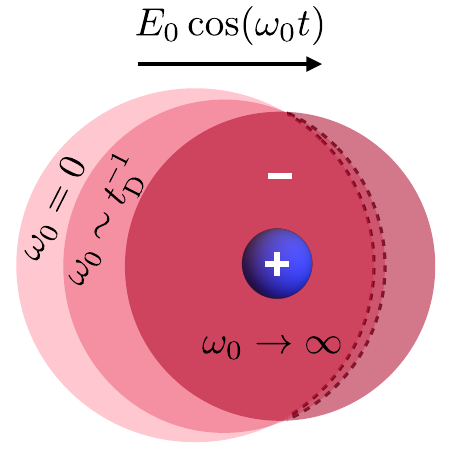}
	\caption{
		Numerically obtained drawing of the negatively charged ionic cloud surrounding a cation in an electrolyte under an oscillatory external uniform electric field, 
		$E_0\cos(\omega_0 t)$, with amplitude $E_0$ and frequency $\omega_0$. The three superimposed red circles depict the distribution of 
		the ionic cloud relative to the cation's position. Their shape is derived from the density-density correlation function between anions and cations 
		[Eq.~(\ref{eq:first_order})] and illustrates the cloud in its most distorted configuration over the oscillation period.}
	\label{fig1}
\end{figure}

One of the earliest investigations into frequency-dependent conductivity was conducted by Debye and Falkenhagen (DF). 
The DF theory~\cite{DebyeFalkenhagen1928} assumes a continuum solvent medium in the low-frequency limit. Moreover, the influence of ion-solvent interactions was not incorporated. 
The DF theory predicts that the real part of the conductivity increases with frequencies in the low-frequency limit~\cite{DebyeFalkenhagen1928}. 
This result can be understood in the following way. The ionic cloud asymmetry exerts a drag force on the moving ion. This asymmetry is established over a timescale of the Debye time, $t\ind{D} \sim \lD^2/D$, where $\lD$ 
is the Debye screening length, and $D$ is the ion diffusion constant. When an AC driving force with a period shorter than the Debye time is applied, the asymmetric 
shape of the ion cloud does not reach its maximal distortion, see Fig.~\ref{fig1}. Hence, the drag force is reduced, and the conductivity increases.

Conversely, a more recent analysis~\cite{anderson1994debye} of the original experimental results by Falkenhagen~\cite{falkenhagen1934} 
has shown that the DF effect has never been adequately validated. 
The challenge in observing this effect is that its magnitude is relatively small, and the model can be justified only at very low concentrations. 
More recently, calculations were carried out for the conductivity at higher concentrations using the {\em mode-coupling theory}, incorporating relaxation of the ionic cloud and hydrodynamic effects~\cite{contreras2013unifying,aburto2013unifying,chandra2000beyond,chandra2000frequency}. These investigations describe the coupling 
between a tagged ion and the collective slow-mode dynamics in terms of several self-consistent equations. It allows a numerical (but not closed-form expressions) evaluation 
of the conductivity in various cases, including high-concentration and time-dependent external fields.  

Advances in nonequilibrium statistical mechanics included the formulation of the {\em stochastic density functional theory} (SDFT)~\cite{Dean1996,kawasaki1994stochastic},
and led to a different approach to calculating the ionic conductivity in the dilute limit~\cite{Demery2016Conductivity,Peraud2017}. 
The governing equation in this approach is often called the Kawasaki-Dean equation and does not rely on an ionic cloud or coupling to the system's collective variables. Instead, SDFT 
accounts for the microscopic ion-ion interactions and allows a systematic derivation of macroscopic quantities in the dilute regime. 

Quite recently, SDFT was employed~\cite{Avni2022JCP,Avni2022PRL} with a modified interaction potential between the ions, taking into account 
the short-range steric interaction. This modified electrostatic potential allowed a better treatment than DHO at a higher concentration range, in which the ions, on average, 
are much closer to one another. The theory successfully predicted the conductivity of NaCl solutions (and other monovalent salts) up to a few molars
without any fit parameters; the DHO theory fails above 10\,mM. Moreover, this framework and the modified potential were used to predict viscosity 
corrections in concentrated electrolytes~\cite{robin2024correlation}.
However, other studies~\cite{bernard2023analytical}
emphasize that, despite the advantages of SDFT coupled with a modified potential, it should be handled
with caution above a few hundred millimolars.

In separate studies, several dynamic aspects of electrolytes have been investigated using SDFT~\cite{Mahdisoltani2021Transient,bonneau2023temporal}. 
It was found that the current response of an electrolyte to a quench of the electric field from finite values to zero is rather complex. 
The main finding is a non-monotonic response followed by an algebraic relaxation of the conductivity as a function of time.

In this study, we employ SDFT to compute the frequency-dependent conductivity of binary monovalent electrolytes. We present the conductivity as a function 
of the interaction potential between the same and oppositely charged ions. We follow Refs. \cite{Avni2022JCP,Avni2022PRL}  and consider similarly the modified 
ion-ion interaction potential that takes into account approximately the ionic hard-core repulsion. Using such modified potential, we compute a closed-form expression 
for the electrolyte conductivity at finite frequencies of an AC external electric field and for an extensive range of concentrations. We show that our theory reduces to the 
DF one without short-range steric interactions. Finally, we show that our theory is relatively robust to different choices of short-range repulsion. 

The outline of this paper is as follows. In Sec.~\ref{sec:model}, we present the model and derive the general expressions for the conductivity of an ionic solution 
with an arbitrary number of species for any interaction potential and time-dependent applied field. In Sec.~\ref{sec:results}, we focus on binary electrolytes 
and derive the frequency-dependent conductivity for different interaction potentials under the weak amplitude limit of the electric field. In Sec.~\ref{sec:Discussion}, 
we discuss our results and their connection to experiments and simulations. Finally, in Sec.~\ref{sec:conclusions}, we conclude and suggest future experiments to test our predictions.

\section{Model}\label{sec:model}

\subsection{Equations of motion}
We consider a system composed of a continuous and homogeneous solvent ({\it e.g.}, water) with dielectric constant $\varepsilon=\varepsilon_r \varepsilon_0$, 
where $\varepsilon_0$ is the vacuum permittivity in SI units and $\varepsilon_r$ is the relative (dimensionless) dielectric constant. Besides the dielectric constant, 
the solvent is characterized by its viscosity $\eta$, and the system is kept at temperature $T$. Cations and anions are modeled as charged Brownian particles,  
solubilized in a three-dimensional fluid, forming an electrolyte solution. The system is then subjected to a time-dependent external electric field that is uniform 
in space, of the form $\boldsymbol{E}(t)=E_0g(t)\hat{\boldsymbol{x}}$, where $\hat{\boldsymbol{x}}$ is the unit vector along the $x$-axis and $E_0$ is the  
electric field amplitude. The function $g(t)$  is a dimensionless function of order unity that encodes the time dependence of the external electric field and has a Fourier transform $\hat{g}(\omega)$. 

Each ionic species $\alpha=1,...,M$, has a charge $z_\alpha e$, where $z_\alpha$ is the valency and $e$ is the electronic unit charge, and its bulk density is $n_\alpha^0$. 
The external field induces a charge current density, $J( E)$, along the same direction as the external field $\EE$. Finally, the DC conductivity is defined by the ratio,
\be
\label{eq:kappa_DC}
\kappa =\left. \frac{J(E)}{E_0}\right|_{E_0=0} \ \ \text{with} \ \ g(t)=1.
\ee

At infinite dilution ($n_{\alpha}^0\to 0$) and constant external field $\EE=E_0 \hat{x}$, the ions perform a Brownian motion with mean velocity 
$e z_\alpha \mu_{\alpha} E_0$, along the field direction, where $\mu_\alpha$ is the ion mobility related to $D_\alpha$ 
by the Einstein relation $\mu_\alpha= D_\alpha/k_{\rm B}T$, and $k_{\rm B}$ is the Boltzmann constant.
The conductivity $\kappa_0$ in the infinite dilution limit, $n_\alpha^{0}\to 0$
is called the Nernst-Einstein conductivity, and is given by $\kappa_0 = e^2\sum_\alpha z_\alpha^2 \mu_{\alpha} n_\alpha^0$. 

At low ionic densities, the interactions between the ions reduce the conductivity. The zero-frequency correction to $\kappa_0$, to leading order in $n_\alpha^0$, is 
given by the DHO result~\cite{onsager1926}. For a binary monovalent electrolyte, it is
\be
\label{eq:DHO}
\kappa\left(\omega{=}0\right) = \kappa_0 \left[1-\frac{r_s}{\lD } -\frac{1}{3}\left(1-\frac{1}{\sqrt{2}}\right)\frac{l\ind{B}}{\lD}\right],
\ee
where $r_s=(6\pi\eta \mu)^{-1}$ 
is the Stokes' hydrodynamic radius of charged particles, 
\be
\lD = \sqrt{\epsilon k\ind{B} T/  2 n_\alpha^0 e^2 }
\ee
is the Debye length,
and $\lB=e^2/(4\pi \varepsilon k_{\rm B}T)$ is the Bjerrum length. At room temperature in water, $l\ind{B}\simeq7$\,\AA.
In the dilute limit, the characteristic length for which the electrostatic interactions are screened is the Debye length, $\lD$. 
Together with the diffusion coefficient averaged over the species types, $\bar{D}$, 
a time scale (called the Debye time) can be constructed,
\be
t\ind{D}= \lD^2/\bar{D},
\ee
and it describes the characteristic relaxation time of the ionic cloud.

Similarly to the definition of the DC conductivity of Eq.~(\ref{eq:kappa_DC}), the frequency-dependent conductivity 
is defined as
\be
\kappa(\omega)=\frac{\hat{J}(\omega, E)}{E_0\hat{g}(\omega)}\bigg\lvert _{E_0=0}\ ,
\ee
where $\hat{J}(\omega,E)$ is the Fourier transform of $J(t,E)$. 

In the following, we study the dynamics of the number density current $\jj_\alpha$ of the ionic species $\alpha$. 
These currents are related to the total average ionic current by
\be
\boldsymbol{J}= e\sum_{\alpha=1}^{M} z_\alpha \langle \boldsymbol{j}_\alpha \rangle
\label{eq:total_current},
\ee
where $\langle ... \rangle$ is a thermal average over the fluctuations. 
We proceed by describing the temporal evolution of the ionic density field $n_\alpha(\rr,t)$ of the species $\alpha$ using SDFT~\cite{Dean1996,Demery2016Conductivity} 
that includes also the hydrodynamic interactions~\cite{Peraud2017,Donev2019,Avni2022PRL}
\bea
	\frac{\partial n_\alpha}{\partial t} &=& - \boldsymbol{\nabla}{\cdot} \boldsymbol{j}_\alpha,  \nonumber\\
	\jj_\alpha
	&=& n_\alpha \boldsymbol{u}  - D_\alpha \boldsymbol{\nabla} n_\alpha
	+ \mu_\alpha   \boldsymbol{f}_\alpha
	+\sqrt{2 D_\alpha n_\alpha}\, \boldsymbol{\zeta}_\alpha.
	 \label{eq:particle_current}
\eea
The first equation is the continuity equation, while the second describes the fluctuating dynamics of the ionic number current density, 
$\jj_\alpha$. The first and second terms of $\jj_\alpha$ in Eq.~(\ref{eq:particle_current}) are the advection and diffusion terms, respectively, 
where $\boldsymbol{u}(\rr,t)$ is the solvent velocity field. The third term accounts for the motion due to the external field and inter-ionic forces, 
where $\boldsymbol{f}_\alpha(\rr,t)$ is the force density acting on the $\alpha$ species particles. Finally, the fourth term in Eq.~(\ref{eq:particle_current}) is a stochastic field, 
where $\boldsymbol{\zeta}(\rr,t)$ is a three-dimensional Gaussian white noise satisfying
\bea
             \langle \boldsymbol{\zeta}_\alpha(\rr,t)\rangle& =& 0, \nonumber\\
             	\langle {\zeta}_\alpha^n(\rr,t) \zeta_\beta^m(\rr',t') \rangle
	&=&  \delta_{\alpha\beta}\delta_{n m} \delta(\rr-\rr')\delta(t-t'),
\label{eq:noise}
\eea
where  $\zeta_\alpha^n$ and $\zeta_\alpha^m$ denotes the $n$-{th} and $m$-{th} cartesian coordinate of the vector $\boldsymbol{\zeta}_\alpha$, $\delta_{ij}$ 
is the Kronecker delta function, and the $\delta$-function is the Dirac delta function in the appropriate dimension.
Throughout this paper, we use the It\^o convention~\cite{Oksendal2000,Dean1996} for the multiplicative noise [as in Eq.~(\ref{eq:particle_current})]. 

The force density $\ff_\alpha$ is the sum of the forces exerted by the external field and force density due to pair interactions with the other ions,
\be
\ff_\alpha=n_\alpha z_\alpha e \boldsymbol{E} -n_\alpha \sum_{\beta=1}^{M} \boldsymbol{\nabla} V_{\alpha\beta}*n_\beta,
\ee
where $V_{\alpha\beta}(\rr)$ is the interaction potential between the $\alpha$ and $\beta$ species, and the symbol $*$ denotes the convolution operator, $h*g \equiv \int \D^3r' \, h(\rr')g(\rr-\rr')$.

The small ionic size typically results in a very low Reynolds number,
meaning that hydrodynamic effects in electrolyte systems are adequately described by
incompressible laminar flow \cite{onsager1926}. Therefore, we assume that the 
fluid velocity field $\uu(\rr,t)$ satisfies the Stokes' equation for incompressible fluids
\bea
	\boldsymbol{\nabla}{\cdot }\boldsymbol{u} & = &0,  \nonumber\\
	 \eta\boldsymbol{\nabla}^2 \boldsymbol{u} - \boldsymbol{\nabla} p &=& -\sum_\alpha  \ff_\alpha, 
	\label{eq:stokes}
\eea
where $p$ is the fluid pressure field.

As we focus hereafter on the frequency-dependent response, our approach neglects inertial effects corresponding to the relaxation of the ion velocity. 
This occurs on a time scale $t\ind{I} ={m\ind{ion} D}/{k\ind{B}T}$, with $m\ind{ion}$ being the ion mass. For frequencies larger than 
$1/t\ind{I}\sim 10^{13}\,\text{s}^{-1}=10$\,THz, the overdamped dynamics description is insufficient. 
Instead, it would be necessary to consider the full Newtonian description 
of the ions and the solvent, as in the Navier-Stokes equation. In addition, our model neglects the frequency dependence of the solvent permittivity. 
This dependence significantly reduces the permittivity for frequencies larger than 10\,GHz \cite{buchner1999,pireddu2024impedance}. 
 Note that we discard the noise term in the Stokes' equation, as this term is divergence-free and will not affect the species density $n_\alpha$, (see, {\em e.g.}, 
 Ref.~(\cite{bonneau2023temporal} for further discussion.)
  
We can integrate over the solvent degrees of freedom $\uu(\rr)$ to obtain a closed-form expression for the current densities, $\jj_\alpha$. 
The solution to Eq.~(\ref{eq:stokes}) is given by \cite{kim2013microhydrodynamics}, 
\be
\uu = \sum_\beta \mathcal{O}*  \boldsymbol{f}_\beta,
\ee
and 
\be
\mathcal{O}_{ij}(\rr) = \frac{1}{8\pi \eta }
\left( \frac{\delta_{ij}}{r}+\frac{r_i r_j}{r^3}\right), \ \ i,j=1,2,3.
\ee
Substituting this result in the expression for the density current, Eq.~(\ref{eq:particle_current}), we get
\be
	\boldsymbol{j}_{\alpha}= -D_{\alpha} \boldsymbol{\nabla} n_{\alpha} +\mu_{\alpha}  \boldsymbol{f}_{\alpha} 
	+n_{\alpha} \sum_\beta \mathcal{O}*  \boldsymbol{f}_\beta 
	+\sqrt{2 D_{\alpha} n_{\alpha}} \boldsymbol{\zeta}_{\alpha}.
	\label{eq:current_expanded}
\ee  

\subsection{Conductivity calculation}
Substituting Eq.~(\ref{eq:current_expanded}) in Eq.~(\ref{eq:total_current}) leads to
\bea
	\JJ &=& \kappa_0 \EE - \sum_{\alpha,\beta} e z_\alpha\mu_\alpha \langle n_\alpha \boldsymbol{\nabla} V_{\alpha\beta}*n_\beta\rangle \nonumber\\
	&+& \sum_{\alpha,\beta} e^2 z_\alpha z_\beta \langle n_\alpha\mcO*n_\beta \rangle\EE .
\eea

Note that the stochastic (noise) term in Eq.~(\ref{eq:current_expanded}) cancels as it is uncorrelated with the density fields at time $t$. 
Moreover, the gradient term cancels as the system is assumed to be homogeneous on average. Lastly, the fourth term of cubic order in the fluctuation
can be discarded as it does not contribute to the linearized theory (see Ref.~\cite{bonneau2023temporal} for further discussion).

Now, we introduce the density fluctuation $\delta n_\alpha(\rr,t)$, for each $\alpha$ species
\be
	\delta n_\alpha(\rr,t)= n_\alpha(\rr,t) - n_\alpha^0 
\ee
 and express the average ionic current $\JJ$ as a function of the density-density correlation given by
\be
		C_{\alpha\beta}(\rr{-}\rr',t) \equiv \langle \delta n_\alpha(\rr,t) \delta n_\beta(\rr',t)\rangle.
	\label{eq:def_correlation}
\ee
Writing the convolution explicitly and using the correlation as defined in Eq.~(\ref{eq:def_correlation}), we find
\bea
	\JJ &= &\kappa_0 \EE 
	\,- \,  \sum_{\alpha,\beta} e z_\alpha\mu_\alpha \int \D^3r \,\boldsymbol{\nabla} V_{\alpha\beta}(\rr)C_{\alpha\beta}(\rr)\nonumber\\
	&+& \sum_{\alpha,\beta} e^2z_\alpha z_\beta \EE \int \D^3r\, \mcO(\rr)C_{\alpha\beta}(\rr).
	\label{eq:current_correlations}
\eea

The correction to the bare current, $\kappa_0\EE$, is the sum of two contributions. The first term involves the electrostatic potential $V_{\alpha\beta}$,
and is referred to 
as the \emph{electrostatic correction} (also known as the relaxation correction). It represents the deformation of the counterion 
cloud around each of the ions by the external electric field. The second term involves the Oseen tensor  $\mcO$ and the external field $\EE$. 
It is called the \emph{hydrodynamic correction} (also known as the electrophoretic correction). It contains the effect of the flow created 
by the counterion cloud under the action of the external field.

The diagonal part of the correlation matrix contains the self-interaction term related to the diffusion constant. Our theory treats diffusion and hydrodynamic 
interaction separately, so to compute the hydrodynamic correction correctly, it is necessary to subtract the self-interaction. This is done by substituting
\be
	C_{\alpha\beta}(\rr{-}\rr',t) \to C_{\alpha\beta}(\rr{-}\rr',t) -n^0_\alpha\delta_{\alpha\beta}\delta(\rr{-}\rr').	
\ee
We remark that this modified correlation does not affect the electrostatic correction. 

Using Parseval's theorem, we express the above equation in Fourier space and obtain
\bea
	\JJ(t)&=& \kappa_0 \boldsymbol{E}(t)+ \sum_{\alpha,\beta} 
	e z_\alpha \mu_\alpha \int \frac{\D^3 k}{(2\pi)^3}  (i\kk) \tilde{V}_{\alpha\beta}(\kk)\tilde{C}_{\alpha\beta}(\kk,t)  \nonumber\\
	&+ &\sum_{\alpha,\beta} e^2 z_\alpha z_\beta \boldsymbol{E}(t) \int \frac{\D^3  k}{(2\pi)^3}\,\tilde{\mathcal{O}}(\kk) 
	\tilde{C}_{\alpha\beta}(\kk,t) \ ,
	\label{eq:avg_J_k}
\eea
where we denote the spatial Fourier transform of $h(\rr)$ as $\tilde{h}(\kk)\equiv\int \dd^3 r\, h(\rr) {\rm e}^{- i \kk \cdot\rr}$. 
Note that the Fourier transform of the interaction potential and Oseen tensor is even in 
$\kk$: $\tilde{V}_{\alpha\beta}(\kk)=\tilde{V}_{\alpha\beta}(-\kk)$ and $\tilde{\mcO}(\kk)=\tilde{\mcO}(-\kk)$.

As we are interested in the frequency-dependent conductivity, we perform an additional Fourier transform from the time domain, $t$, into the frequency domain, 
$\omega$, 
\bea
\hat{\JJ}(\omega) &=&  \kappa_0\hat{\boldsymbol{E}}(\omega)+ \sum_{\alpha,\beta} 
e z_\alpha \mu_\alpha \int \frac{\D^3 k}{(2\pi)^3}\,  (i\kk) \tilde{V}_{\alpha\beta}(\kk)\hat{C}_{\alpha\beta}(\kk,\omega)  \nonumber\\
	&+& \sum_{\alpha,\beta} e^2 z_\alpha z_\beta  \int \frac{\D^3 k }{(2\pi)^3}   \nonumber \\ 
	& &\times \int \D \omega' \, 
	\tilde{\mathcal{O}}(\kk) \hat{C}_{\alpha\beta}(\kk,\omega') \hat{\boldsymbol{E}}(\omega-\omega')  \, ,
	\label{eq:avg_J_k_omega}
\eea
where  the spatio-temporal Fourier transform of a function $h(\rr,t)$ is denoted as $\hat{h}(\kk,\omega)\equiv\int \dd^3 r\, 
\dd t \, h(\rr,t) {\rm e}^{- i \kk \cdot\rr-i \omega t}$. 
The above Eq.~(\ref{eq:avg_J_k_omega}) is the general expression for the Fourier transform of the current $\hat{\JJ}(\omega)$ expressed in terms 
of the Fourier transform of any two-body potential $\tilde{V}_{\alpha\beta}(\kk)$, external electric field $\hat{\boldsymbol{E}}(\omega)$ and 
density-density correlations $\hat{C}_{\alpha\beta}(\kk,\omega)$. Finally, note that in our study, $J$ and $E$ are temporal functions that are homogeneous in space.

\subsection{Linearized SDFT}
To get more straightforward analytical results, we return to Eq.~(\ref{eq:current_expanded}) in position space and 
linearize it around the ionic bulk densities, $ n_{ \alpha}^0$. The resulting evolution equation for the ionic density fluctuations 
of the $\alpha$ species, $\delta n_\alpha=n_\alpha -  n_{ \alpha}^0$, reads,
\bea
	\frac{\partial }{\partial t} {\delta n}_\alpha &=& D_{\alpha}\boldsymbol{\nabla}^2 \delta n_{\alpha} -  
	\mu_{\alpha}e z_\alpha \boldsymbol{E}(t)\cdot \boldsymbol{\nabla} \delta n_\alpha  \nonumber\\ 
	&+& \mu_{\alpha}n_\alpha^0 \boldsymbol{\nabla}^2\left[\sum_\beta V_{\alpha\beta}*\delta n_\beta\right]  
	+\sqrt{2D_{\alpha}n_{\alpha}^0}~ \boldsymbol{\nabla}{\cdot}\zzeta_{\alpha} \  .\nonumber\\ 
	\label{eq:linear_sdft_real_space}
\eea
The dynamic equation can be written in simplified form in Fourier $k$-space
\be
	\frac{\dd }{\dd t}{\delta \tilde{  n}_\alpha} = -  A_{\alpha\beta} \delta \tilde{  n}_\beta + \chi_\alpha,
	\label{eq:vectorial}
\ee
where the matrix $\mathbf{A}$ is given by
\be
	A_{\alpha\beta}(\kk,t) = \delta_{\alpha\beta} \left[ D_\alpha k^2 + 
	i \mu_{\alpha} e z_\alpha E(t) k_x \right] +  \mu_{\alpha} n_\alpha^0 k^2 \tilde{V}_{\alpha\beta}.
\ee

We also introduced a scalar Gaussian noise $\chi_\alpha(\kk,t)$,
that satisfies
\be
\langle\chi_\alpha(\kk,t)\chi_\beta(\kk',t')\rangle = 2 (2\pi)^3 B_{\alpha\beta} \, \delta(\kk+\kk')\delta(t-t'),
\ee
with a diagonal noise correlation matrix  $B_{\alpha\beta} =  \delta_{\alpha\beta} n_\alpha^0 D_\alpha k^2 $. 
Hence, in Fourier space, the density-density correlations are given by
\be
\label{eq:correlation_in_fourier}
	\langle \delta \tilde{n}_\alpha(\kk,t) \delta \tilde{n}_\beta(\kk',t)\rangle= 
	(2\pi)^3 \delta(\kk+\kk')\tilde{C}_{\alpha\beta} (\kk,t).
\ee
Using the It\^o product rule (see Ref.~\cite{Gardiner2009}, Secs. 4.3 and 4.4) in Eq.~(\ref{eq:vectorial}), we find that the correlation matrix $\tilde C$ satisfies
\be
	\frac{\D}{\D t} \tilde{ \mathbf{C}}
	= 2\mathbf{B} -\mathbf{A}\tilde{\mathbf{C}}- \tilde{\mathbf{C}} \mathbf{A}^{\text{*}},
	\label{eq:correlation_1}
\ee
where the matrix $\mathbf{A}^{\text{*}}$ is the complex conjugate of $\mathbf{A}$. Equation~(\ref{eq:correlation_1}) 
is a set of linear inhomogeneous (with a source) and non-autonomous (with an explicit dependence on the time variable $t$), 
first-order ordinary differential equations (ODEs). In the next section, we consider binary monovalent electrolytes for which Eq.~(\ref{eq:correlation_1}) can be further simplified.

\section{Results}\label{sec:results}

\subsection{Binary monovalent electrolyte}
%
For binary monovalent 1:1 electrolytes (such as table salt, NaCl), the electrolyte is composed of cation-anion pairs, $\alpha=\pm $, 
with charges $z_\pm e=\pm e$. Because of overall charge neutrality, $ n^0_\pm=n_0$, we further assume that the ionic mobilities are the same, $\mu_\pm=\mu$, implying $D_{\pm}=D$.

Using the symmetry of the 1:1 electrolytes, we know that $\hat{C}_{--}= \hat{C}_{++}$ and  $\hat{C}_{+-}= \hat{C}^\text{*}_{-+}$. 
Hence, the elements  of the $2\times 2$ correlation matrix can be expressed via three independent functions, $\hat{c}$, $\hat{r}$ and $\hat{m}$,
\bea
\hat c(\kk,\omega) &\equiv&\frac{1}{n_0} \hat{C}_{++}(\kk,\omega) \, ,\nonumber\\
\hat r(\kk,\omega) &\equiv& \frac{1}{n_0}\text{Re}\left[\hat{C}_{+-}(\kk,\omega)\right] \, ,\nonumber\\
\hat m(\kk,\omega) &\equiv& \frac{1}{n_0}\text{Im}\left[\hat{C}_{+-}(\kk,\omega)\right]  \, .
\eea

We then choose the following form of the modified electrostatic potential in $k$-space, $\tilde{V}_{\alpha\beta} = (e^2/\varepsilon ) v_{\alpha\beta}(k, a)$ 
where $k\equiv|\bf{k}|$ and $a$ is a length scale related to the short-range repulsion between the ions. The behavior of $v_{\alpha\beta}$ at $k a \ll1 $ 
recovers the standard Coulomb potential, $\sim 1/k^2$, but deviates when $k a \approx 1$, where its functional form depends on $a$.

It is convenient to introduce dimensionless variables. The spatial coordinates are rescaled by the Debye length so that $\qq \equiv \kk\lD$, while  
time and frequency are rescaled 
by the Debye time so that $\Omega\equiv\omega t\ind{D}$ and $\tau \equiv t / t\ind{D}$. Applying these dimensionless variables, 
the integrals in Eq.~(\ref{eq:avg_J_k_omega}) are rewritten in a dimensionless form. By dividing $\hat{J}(\Omega,E)$ by $ \hat{g}(\Omega)E_0$, 
one gets the frequency-dependent conductivity
\be
\frac{\hat{J}(\Omega,E)}{ \hat{g}(\Omega)E_0 } =\kappa_0+\kappa \ind{el}(\Omega)+\kappa \ind{hyd}(\Omega),
\label{eq:corrections}
\ee
where the electrostatic and hydrodynamic corrections read, respectively
\bea
	\kappa\ind{el}(\Omega,\mathcal{E}  ) &=&  -\frac{\kappa_0}{2 \pi^2}  \frac{\lB }{\lD} \frac{1}{\hat{g}(\Omega)\mathcal{E}}
	 \int  \D^3 q\, q_x \hat m(\Omega,q) v_{_{+-}}(q)\,, \nonumber\\
	 	& & \nonumber \\
	\kappa\ind{hyd}(\Omega,\mathcal{E}) &=& \frac{3\kappa_0}{4 \pi^2 } \frac{r_s}{\lD} 
	\frac{1}{\hat{g}(\Omega) } \int \D \Omega' \int \frac{\D^3 q} {q^2}\big( 1{-} q_x^2/q^2\big) \nonumber \\ 
	 & &\times \Big[ \hat c(\Omega',q){-}
	\hat r(\Omega',q){-}\delta(\Omega')\Big]  
	  \hat{g}(\Omega{-}\Omega') \, , 
	 \label{eq:gamma_hyd}
\eea
where $\mathcal{E} \equiv e \lD E_0/k_{\rm B}T$ is the dimensionless electric field. Using the same rescaling, we can write the SDFT equation for the correlation functions as
\bea
		\frac{\D \tilde c}{\D \tau}  &= &  - q^2  \left(2+   v_{++}\right) \tilde c(\tau)- q^2v_{-+}     \tilde r(\tau)+2 q^2, \nonumber \\
		\frac{\D \tilde r}{\D \tau}  &=&    -q^2  \left(2+   v_{++}\right)\tilde r(\tau) - q^2v_{-+}   \tilde c(\tau)+2 \mathcal{E}  q_x g(\tau)\tilde m(\tau), \nonumber  \\
		\frac{\D \tilde m}{\D \tau} &=& -q^2  \left(2+   v_{++}\right)\tilde m(\tau)-2 \mathcal{E}  q_x g(\tau) \tilde r(\tau). 
	\label{eq:dimensionless_sdft}
\eea
Finding a general closed-form solution to the set of Eq.~(\ref{eq:dimensionless_sdft}) is difficult. Hence, we look for periodic solutions in the small electric field 
$\mathcal{E}$ limit. To do so, we examine the temporal Fourier transform of  Eq.~(\ref{eq:dimensionless_sdft}) 
\bea
		\hat c(\Omega) &=& \frac{2 q^2\delta(\Omega)- q^2v_{_{-+}} \hat r(\Omega)}{ \Lambda(\Omega) } \, , \nonumber \\
		\hat r(\Omega) &=& \frac{  
		2\mathcal{E} q_x  \left(\hat m * \hat{g}\right)(\Omega) - q^2v_{_{-+}}    \hat c(\Omega)}{\Lambda(\Omega)} \, , \nonumber \\
		\hat m(\Omega) &=& -\frac{2\mathcal{E} q_x \left(\hat r * \hat{g}\right)(\Omega) }{\Lambda(\Omega)} \, . 
		\label{eq:corr_c_r_m}
\eea
where $\Lambda(\Omega) = i\Omega+q^2  \left(2+ v_{_{++}}\right)$. 

\begin{figure*}
	\centering
	\includegraphics[width=0.8\linewidth]{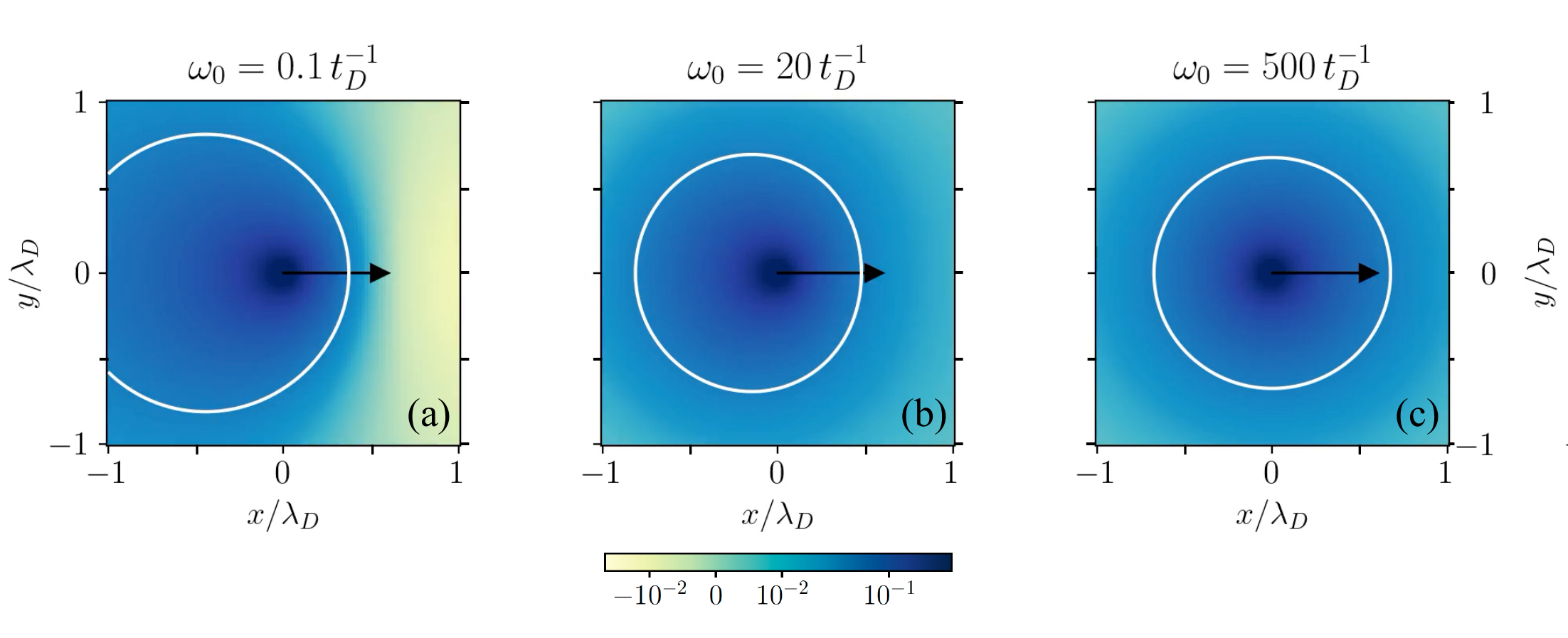}
	\caption{The density-density correlation function $c_{+-}(x/\lD,y/\lD,t)$, where the lengths are measured in units of $\lD$. The correlation $c_{+-}$ 
	is computed by taking numerically the Fourier transform of the solution in Eq.~(\ref{eq:first_order}), for $g(t) = \cos(\omega_0 t)$. A pure Coulomb 
	interaction potential is used. The three panels show a color map of $c_{+-}$ at $t=0$ (when the electric field magnitude is maximal) 
	for different values of $\omega_0$. In order to enhance the figure's deformation visibility, we set $\mathcal{E}=3$. The arrows indicate 
	the direction of the electric field. In panel (a), the oscillation is slow compared to $t\ind{D}$. Therefore, the correlation shape reaches 
	a strongly asymmetric form along the direction of the electric field. Panel (c) shows the opposite case, when the field oscillation is fast compared to $t\ind{D}$, 
	leading to an almost spherically symmetric correlation. Panel (b) shows an intermediate case.
	See [link] for an animation of the correlation function evolution over time.
	}	
	\label{fig2}
\end{figure*}

Solving these equations that give the system's limit cycle is not straightforward. Yet, one can solve them to leading order in $\mathcal{E}$ 
and get the frequency-dependent conductivity computed at vanishing electric fields
\bea\label{eq:first_order}
 \hat c^{(0)}(\Omega)& =& \frac{4+2v_{++}}{\left(2+v_{++}\right)^2-v^2_{-+}}\delta(\Omega) \, , \nonumber\\
\hat r^{(0)}(\Omega)  &=& -\frac{2v_{+-}}{\left(2+v_{_{++}}\right)^2-v^2_{-+}}\delta(\Omega) \, , \\
\hat m^{(1)}(\Omega) &=&  \frac{4 q_x v_{+-} \hat{g}(\Omega) }{\left[\left(2+v_{++}\right)^2-v^2_{-+}\right]\Big[i\Omega+q^2(2+v_{++})\Big]}, \nonumber
\eea
where the superscript $(i)$ denotes the $i$th term in the power expansion of $\mathcal{E}$. Note that the terms $\hat m^{(0)}$, $\hat c^{(1)}$, and $\hat r^{(1)}$ 
are equal to zero and are discarded. In Fig.~\ref{fig2}, the real-space and time-dependent correlation function $c_{+-}(x/\lD,y/\lD,t)$ 
is computed for Coulombic interaction, and for the single mode, $g(t) = \cos(\omega_0 t)$. 

%
Expressing the integrals in Eq.~(\ref{eq:gamma_hyd}) using Eq.~(\ref{eq:first_order})
yields
\bea
	\kappa\ind{el}(\Omega) &=&  -\frac{8\kappa_0}{3\pi} \frac{\,\lB}{\lD}	
	 \int_0^\infty  \dd q \nonumber \\ 
	 & &\times \frac{q^4 v^2_{+-}}{\left[\left(2+v_{_{++}}\right)^2-v^2_{+-}\right]\Big[q^2\left(2+v_{_{++}}\right)
	 +i\Omega\Big]} \, ,
	 \nonumber\\ 
	 	& & \nonumber \\
	\kappa\ind{hyd} &=& -\frac{2\kappa_0}{\pi} \frac{\,r_s}{\lD} 
	 \int_0^\infty  \dd q \, \frac{v_{_{++}}-v_{_{+-}}}{2+v_{_{++}}-v_{_{+-}}}  \, . 
  \label{eq:conductivity with v2}
\eea
Equation~(\ref{eq:conductivity with v2}) gives the frequency-dependent 
conductivity for a binary monovalent electrolyte as a function of the interaction potential between the ions.

\subsection{Pure Coulomb potential}
Evaluating the integrals in Eq.~(\ref{eq:conductivity with v2}) for Coulomb potential, namely 
\be
v_{_{++}}=-v_{_{+-}} = 1/k^2,
\ee
recovers the well-known Debye-Falkenhagen (DF) result~\cite{DebyeFalkenhagen1928}  
\be
	\kappa\ind{el}\left(\omega\right)=-\frac{\kappa_0}{3 }
	\frac{\lB}{\lD}\frac{1}{2+\sqrt{2+2 i \omega t\ind{D}}}\, .
	\label{eq:el correction coulomb}
\ee
where the DF result is expressed in terms of the physical frequency, $\omega=\Omega t\ind{D}$. 
Examining Eqs.~(\ref{eq:gamma_hyd}) and (\ref{eq:first_order}), we see that the hydrodynamic correction 
depends on the zeroth order terms in $\mathcal{E}$ of the correlations. Thus, it is independent of $\omega$,
\be
	\kappa\ind{hyd} = -\kappa_0 \frac{r_s}{\lD} \, .
		\label{eq:hyd correction coulomb}
\ee
We can also see that the standard DHO correction is recovered at the $\omega\to0$ limit [Eq.~(\ref{eq:DHO})]. 
Similarly to the DHO result, Eqs.~(\ref{eq:el correction coulomb}) and (\ref{eq:hyd correction coulomb}) are limited to small concentrations, typically lower than $10$\,mM.

\subsection{Modified potentials}
The original DF result can be improved by introducing an interaction potential that accounts for the repulsion between the ions at short ranges.

An effective potential that takes into account the steric interaction should contain the scale for which the steric repulsion between ions becomes substantial, 
namely, when the distance between the two particle centers is equal to the sum of their radii $r_\alpha+r_\beta$.  For simplicity, we define $a$ as such a 
characteristic length, ignoring the size difference between ions of different species. 
  
The truncated Coulomb potential has been previously proposed to account for the short-range interaction between ions~\cite{adar2019,Avni2022PRL,Avni2022JCP}. It is written as
\be
V_{\alpha\beta}(r) = \frac{e^2 z_\alpha z_\beta}{4\pi \varepsilon }\frac{\theta(r-a)}{r} \, ,
\ee
where $\theta$ is the Heaviside function. Using the truncated potential in Eq.~(\ref{eq:conductivity with v2}) gives the following expressions for the two 
conductivity correction terms
\bea\label{eq:kappa_truncated}
	\kappa\ind{el}(\omega) &=&  -
	\frac{2\kappa_0}{3\pi }\frac{  \lB }{ \lD  } \int_0^\infty \D q\, \nonumber \\ 
	& &\times
	 \frac{q^2 \cos ^2 \left(\frac{a q}{\lD}\right)}{ \Big[\cos \left(\frac{a q}{\lD}\right)
	+ q^2 \Big] \Big[\cos  \left(\frac{a q}{\lD}\right) + 2q^2+i \omega t\ind{D} \Big]}    \, , \nonumber \\
	& & \nonumber \\
	\kappa\ind{hyd} &=& -\frac{2\kappa_0}{\pi} \frac{ r_s}{\lD}  
	 \int_0^\infty  \dd q \frac{\cos \left(\frac{a q}{\lD}\right)}{\cos  \left(\frac{a q}{\lD}\right)+q^2} . 
\eea

Another modified potential is a soft truncated potential that was proposed in 
Refs.~\cite{santangelo2006computing,lee2015dynamics,desouza2020continuum}
\be
\label{eq:stp}
V_{\alpha\beta}(r) =  \frac{e^2 z_\alpha z_\beta}{4\pi \varepsilon }\left(\frac{1}{r}-\frac{{\rm e}^{-r/a}}{r}\right) \, .
\ee
This potential gives simpler polynomial expressions in Fourier space that can be integrated exactly. However, the exact expressions 
are rather elaborate and will not be presented here. Substituting Eq.~(\ref{eq:stp}) in Eq.~(\ref{eq:conductivity with v2}), we get
\bea \label{eq:kappa_stp}
	\kappa\ind{el}(\omega) &= & -\frac{2\kappa_0}{3\pi }\frac{ \lB }{ \lD  }  \int_0^\infty \D q \, \frac{q^2} {\left[\big(\frac{a}{\lD}\big)^2 q^4+q^2+1 \right]} 
	 \nonumber \\ 
	& &\times  \frac{1} { \left[2 \big(\frac{a}{\lD} \big)^2 q^4 
	+ q^2 \left(2+i \frac{a^2}{D} \omega  \right)+i \omega t\ind{D} +1\right]}, \nonumber \\
       & & \nonumber\\
	\kappa\ind{hyd} &= &-\frac{2\kappa_0}{\pi} \frac{r_s}{\lD}  \int_0^\infty  \D q \frac{1}{\big(\frac{a}{\lD}\big)^2 q^4+q^2+1} \, . 
\eea

\begin{figure}
	\centering
	\includegraphics[width=0.9\linewidth]{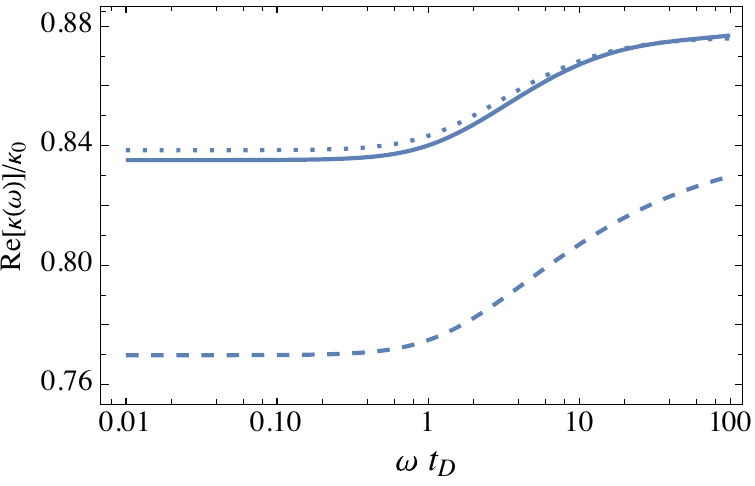}
	\caption{The real part of the normalized conductivity as a function of $\omega t\ind{D}$ for the three interaction potentials. The 
	Coulomb potential [Eqs.~(\ref{eq:el correction coulomb}) 
	and (\ref{eq:hyd correction coulomb})] is plotted with a dashed line, the truncated potential [Eq.~(\ref{eq:kappa_truncated})] 
	with a solid line and the soft    
	truncated potential [Eq.~(\ref{eq:kappa_stp})] with a dotted line. The solution concentration is taken as $n=0.1$\,M.
	The other parameter values are $l_\text{B} = 7$\,\AA, 
	$\lD = 10$\,\AA, $r_\text{s} = 1.5$\,\AA\ and $a = 3$\,\AA.}	
	\label{fig3}
\end{figure}

\section{Discussion}
\label{sec:Discussion}
By examining Eq.~(\ref{eq:conductivity with v2}), we can see that the electrostatic correction $\kappa\ind{el}$ depends on the external-field 
frequency $\omega$. In contrast, the hydrodynamic correction $\kappa\ind{hyd}$ remains the same as in the stationary case. In Fig.~\ref{fig3}, 
we compare the numerical evaluations of $\kappa(\omega)$ [Eq.~(\ref{eq:corrections})] with the three different potentials:
Coulomb, truncated Coulomb, and soft truncated Coulomb. 

For simple ions in aqueous solutions at room temperature, the parameter values used are 
$l_\text{B} = 7$\,\AA, $\lD = 3$\,\AA $/\sqrt{n[{\rm M}]}$, $r_\text{s} = 1.5$\,\AA\ and $a = 3$\,\AA. 
The Debye time has typical value of $t\ind{D} = (10^{-10} /n[{\rm M}]) \, \text{s}$. 
One can see that the conductivity increase starts when $\omega\,  t\ind{D} \simeq 1$. 
The truncated and soft truncated Coulomb potentials give roughly the same conductivity for $a\simeq \lambda\ind{D}$ at moderate frequencies.
Different choice of potentials affects the correction to the conductivity at driving frequencies that exceed $\omega\, t\ind{D}\gtrsim 1000$. 
However, within this range of very high frequencies, our model does not hold as it is based on overdamped dynamics.

\begin{figure}
	\centering
	\includegraphics[width=0.9\linewidth]{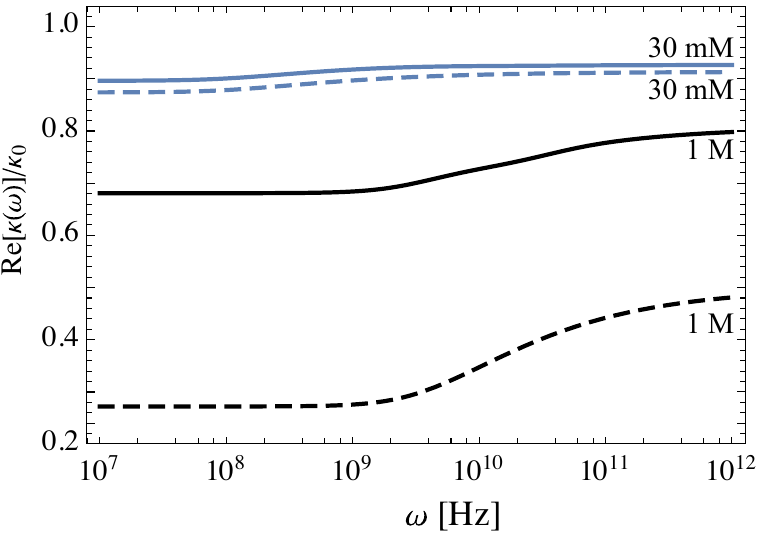}
	\caption{The real part of the normalized conductivity rescaled by $\kappa_0$, for different concentration values as a function 
	of the external field's oscillation frequency, 
	$\omega$. The normalized conductivity is plotted for the truncated potential (solid) and the Coulomb potential (dashed) 
	for concentrations 30\,mM and 1\,M. 
	The frequency value at which the conductivity transitions from a pseudo-plateau that equals the static conductivity to an increasing 
	function scales roughly with $n$. 
	Other parameter values are the same as for Fig.~\ref{fig3}.  
	}	
	\label{fig4}
\end{figure}

The real part of the normalized conductivity is plotted in Fig.~\ref{fig4} for two very different concentration values, 30\,mM and 1\,M,  
as a function of the driving oscillation frequency in Hz. At low frequencies, the conductivity behaves as the static one $(\omega=0)$, 
but as $\omega$ increases, the conductivity increases towards $\kappa=\kappa_0+\kappa\ind{hyd}$,  {\it i.e.}, $\kappa_{\rm el}$ approaches zero. 
One can see that the frequency value at which this changeover occurs scales roughly with $n$. In other words, the critical frequency $\omega\ind{D}$ 
below which the conductivity 
behaves as the static conductivity. We recall that the critical frequency increases linearly with the ionic concentration,  $\omega\ind{D} = 1/t_D \sim n$.

\begin{figure}
	\centering
	\includegraphics[width=0.9\linewidth]{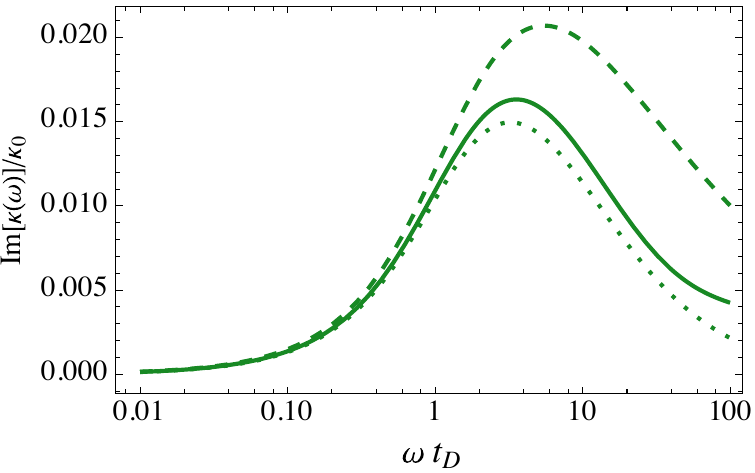}
	\caption{The imaginary part of the normalized conductivity (rescaled by $\kappa_0$) for Coulomb potential (dashed line), truncated potential 
	[solid line -- Eq.~(\ref{eq:kappa_truncated})], and the soft truncated potential [dotted line -- Eq.~(\ref{eq:kappa_stp})] for $n{=}0.1$\,M. 
	For high frequencies $\omega\gg t\ind{D}^{-1}$ and low ones $\omega\ll t\ind{D}^{-1}$, the imaginary part is small, and the charge current is 
	in phase with the external field. Other parameter values are the same as for Fig.~\ref{fig3}. 	
	}	
	\label{fig5}
\end{figure}

In Fig.~\ref{fig5}, the imaginary part of the normalized conductivity is plotted for the three interaction potentials. 
For pure Coulomb potential, the imaginary part accounting for the response phase-shift starts to decay at frequency $\omega\, t\ind{D}\approx10$, 
while for the truncated and soft-truncated potentials, the decay begins at $\omega\, t\ind{D}\approx 2$.
This leads to the conclusion that including the steric interactions in the interaction potential reduces the dephasing effect.
 
\begin{figure}
	\centering
	\includegraphics[width=0.9\linewidth]{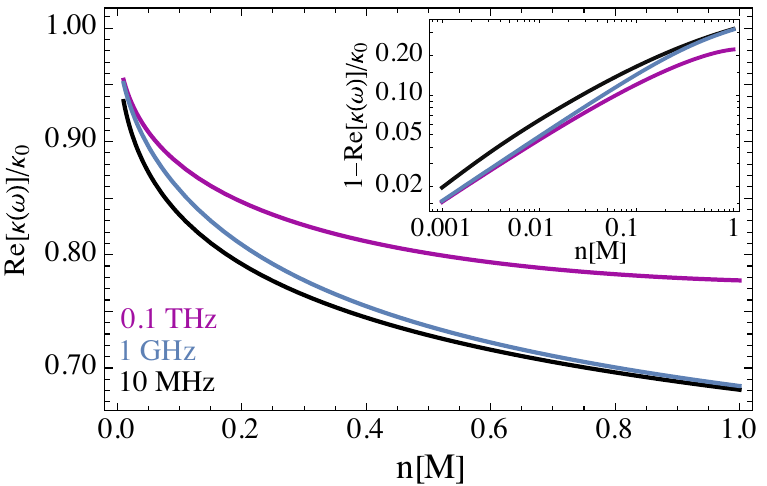}
	\caption{ The real part of the normalized conductivity as a function of the ion concentration for three driving frequencies, 
	$\omega= 10$\,MHz (black line), 1\,GHz (blue line), and 0.1\,THz (purple line), using the truncated potential. 
	The blue curve for the intermediate frequency of 1\,GHz crosses over between the static conductivity behavior at low frequencies
	(black line, $\omega{=}10$\,MHz) and the high-frequency conductivity behavior (purple line, $\omega{=}0.1$\,THz). The crossover 
	is due to the dependence of the Debye time $t\ind{D}$ on the concentration. 
	In the inset, the rescaled (with $\kappa_0$) deviation from the Nernst-Einstein conductivity is shown on a log-log scale, where the cross-over is visible at 
	both ends. Other parameter values are the same as for Fig.~\ref{fig3}.
	}
	\label{fig6}
\end{figure}

Fig.~\ref{fig6} presents the real part of the normalized conductivity as a function of the ion concentration at different driving frequencies, 
$\omega=$ 10\,KHz, 1\,GHz, and 0.1\,THz (using only the truncated potential). One can see a crossover between the static and high-frequency conductivity 
in the blue curve corresponding to the intermediate frequency, $\omega=$ 1\,GHz. The crossover is due to the dependence of 
the Debye time $t\ind{D}\sim 1/n$ on the concentration, leading to a change between static and high-frequency conductivity. 

Our model is based on simple considerations that restrict it from describing all the rich phenomenology of electrolyte systems
self-consistently; in particular, it assumes a constant 
dielectric susceptibility and viscosity. However, these quantities depend on the ionic concentration and the external field, and incorporating 
such refinements into the model is left to future work. 

\paragraph*{Relation to experiments --}
 
The frequency-dependent conductivity was measured in some experiments
\cite{querry1972optical,buchner1999,peyman2007complex,vinh2015high,chikina2023ac}. Unfortunately, there are specific 
difficulties in comparing our predictions with experimental results. 
First, the dielectric susceptibility of water strongly decreases as a function of the dissolved salt concentration. This is the celebrated ``dielectric decrement'' 
phenomenon, observed experimentally and discussed in Refs.~\cite{biesheuvel2005volume,ben2011dielectric,hatlo2012electric,lopez2014influence}.
 
In addition, the dielectric susceptibility decreases at high frequencies of the order of GHz   \cite{buchner1999,peyman2007complex,pireddu2024impedance}. 
This effect alters the electrostatic interaction between the ions. Moreover, the increase in conductivity with the external field frequency
happens at frequencies similar to the resonance frequency of water molecules. This leads to several complications with the measurements, including heating 
of the aqueous solution.

Finally, in high-frequency experimental setups, boundary effects around the electrodes extend well into the electrolyte solution,
making the measurement of bulk conductivity at high frequencies quite challenging~\cite{chikina2023ac,chikina2024role}.
To the best of our knowledge, the frequencies at which the conductivity was measured so far are lower than a few MHz. Up to these frequencies, 
the behavior is quasi-static; the conductivity does not depend on $\omega$ and is equal to its DC values. 
These observations make it hard to validate our predictions with relevant experimental data, although
these predictions have the correct qualitative behavior as a function of salt concentration and driving frequency. 

We also would like to comment on molecular-dynamic (MD) simulations that have been done for similar 
frequency-dependent conductivity  \cite{pireddu2024impedance,stoltz2023frequency,chandra1993frequency}. 
To observe the Debye-Falkenhagen (DF) effect in an MD simulation, it is necessary to eliminate the dipolar effects of 
the solvent, as the change in the dielectric constant would mask the DF effect. 
A similar setting was investigated in Ref.~\cite{stoltz2023frequency}. 
It was shown that the confinement (finite-size effect) has a significant impact on the electrolyte dynamics. 
This renders the bulk behavior challenging to observe. 

\section{Conclusions}
\label{sec:conclusions}
%
We calculated the frequency-dependent conductivity for binary monovalent electrolytes within the stochastic density functional theory (SDFT) framework. 
We examined several modified ion-ion interaction potentials to account for the short-range effect missing in the Debye-Falkenhagen (DF) theory 
that does not include the ion steric effect. By employing two modified Coulomb potentials that suppress unphysical, short-range electrostatic attraction, 
we demonstrate that the correction is relatively robust in terms of the details of the modified potential. 

The DF effect is relatively small and, for various monovalent salts, falls in the same frequency range as water's resonance frequency. This leads to a masking 
of the DF effect by the significant change in the relative dielectric constant. To adequately compare our findings to experiments, one would need to measure 
the conductivity of a 1:1 electrolyte in a solvent that does not present a strong frequency dependence of the dielectric constant in the GHz range. Alternatively, 
simulations that assume an implicit solvent may be used to avoid effects from the water molecules' polarization. 

The theory presented here can be generalized to multi-component electrolytes and multivalent ions. However, the latter is expected to limit the theory's validity 
to lower concentrations due to strong electrostatic correlations and fluctuations~\cite{Avni2022JCP}, which are not fully considered within our work.

\section{Acknowledgments}
The authors thank S. Nakamae, I. Chikina, and A. Varlamov for their useful discussions of experimental works.
DA acknowledges the NSFC-ISF Research Program, jointly funded by the National Science Foundation China (NSFC) and the Israel Science 
Foundation (ISF) under grant No. 3396/19, and ISF grants No. 213/19 and 226/24. 
YA acknowledges support from the Kadanoff-Rice Postdoctoral Fellowship and the Zuckerman STEM Leadership Program.



\end{document}